\documentclass[12pt]{iopart} 

\usepackage{amsthm}
\expandafter\let\csname equation*\endcsname\relax
\expandafter\let\csname endequation*\endcsname\relax
\usepackage{mathptmx}
\usepackage{dsfont}
\usepackage{mathbbol}
\usepackage{bbold}
\usepackage{mathrsfs}
\usepackage{graphicx}
\usepackage{dcolumn}
\usepackage{bm}
\usepackage{comment}
\usepackage{verbatim}
\usepackage[usenames]{color}
\usepackage[normalem]{ulem}
\usepackage{cancel} 

\usepackage[per-mode=symbol]{siunitx}   
\usepackage[version=3]{mhchem}
 
\DeclareSIUnit\intensity{\watt\per\centi\meter\squared}
\DeclareSIUnit\fieldstrength{\volt\per\centi\meter}
\DeclareSIUnit\kfieldstrength{k\volt\per\centi\meter}
\DeclareSIUnit\energy{cm^{-1}}

\newcommand{\melement}[3]{\ensuremath{\left\langle #1 \left|#2\right|#3\right\rangle}}%
\newcommand{\ie}{i.\,e.}%

\newcommand{\geougr}{\address{Departamento de Geometría y Topología, Facultad de Ciencias, Universidad de Granada, 18071 Granada, Spain}}%
\newcommand{\ucm}{\address{Departamento de Qu\'imica F\'isica, Universidad Complutense de Madrid, 28040 Madrid, Spain}}

\begin{document}

\title{Revisiting Quantum Optimal Control Theory: New Insights for the Canonical Solutions}
\author{Katherine Castro}\geougr
\author{Ignacio R. Sol\'a}\ucm
\author{Juan J. Omiste}\ucm\ead{jomiste@ucm.es}
\date{\today}
\begin{abstract}
In this study, we present a revision of the Quantum Optimal Control Theory (QOCT) originally proposed by Rabitz et al~\cite{Peirce1988}, which has broad applications in physical and chemical physics. First, we identify the QOCT equations as the Euler-Lagrange equations of the functional associated to the control scheme. In this framework we prove that the extremal functions found by Rabitz 
are not continuous, as it was claimed in previous works. Indeed, we show that the costate is discontinuous and vanishes after the measurement time. In contrast, we demonstrate that the driving field is continuous. We also identify a new set of continuous solutions to the QOCT. Overall, our work provides a significant contribution to the QOCT theory, promoting a better understanding of the mathematical solutions and offering potential new directions for optimal control strategies.

\end{abstract}

\maketitle

\section{Introduction}
\label{sec:introduction}
Minimization and optimization are standard tools in mathematics that we encounter in many areas of science~\cite{Goldstine1980,Stewart2011, Goldstein2002} ranging from health and biology to space exploration or even social behaviour~\cite{Mesterton-Gibbons2009}. Furthermore, most of the fundamental equations in physics may be derived from a variational principle, which aims to minimize a quantity, for instance, the action or the energy~\cite{Jackson1999,Goldstein2002,Miller2010a}. Prominent examples are Lagrangian mechanics, general relativity or geometric optics. In atomic and molecular physics, let us highlight the use of the so-called Dirac-Frenkel variational principle to propagate quantum systems with many degrees of freedom~\cite{Raab2000}, which leads to the Multiconfigurational Time-Dependent Hartree and related methods~\cite{Meyer1989,Miyagi2013,Beck2000a,Sato2013,Hochstuhl2014}.

In the literature, we find 
Optimal Control as the methodology designed to obtain a driving field such that a given observable is optimal, \ie, a maximum or minimum, for which a functional is designed and its extremal solutions are explored, ensuring that certain constraints are fulfilled~\cite{Bulirsch1993,Brif2010, Sola2018,Poznyak2008,Prados2021}.  
It is said to be Quantum Optimal Control if the Schr\"odinger equation drives the dynamics~\cite{Peirce1988}. This methodology has been widely used to control molecular orientation~\cite{artamonov:phys_rev_a_82_023413, Salomon2005,Coudert2017}, state transitions and population transfer~\cite{Sola1998,Sundermann1999,Wu2004,Basilewitsch2019,Coden2020}, ultracold atoms systems~\cite{Jensen2020a,Jensen2020b}, ionization~\cite{Larsson2020} or even entanglement~\cite{Yu2018a,Magann2019}. It has also been used to drive Bose-Einstein condensates~\cite{Dupont2021} using the classical~\cite{Amri2019} and Gross-Pitaevskii approximation~\cite{Mennemann2015} or to drive dissipative systems~\cite{Bartana1997,Ohtsuki1999,Bonnard2009,Bonnard2009a}.

The most used algorithms are based on the gradient of the functional with respect to the field~\cite{Werschnik2007,BalintKurtiACP08,Goerz2022}, 
which can be used to ensure a fast monotonic increase in the objective.
This is the case of Krotov's method~\cite{Krotov1983,Tannor1992,Somloi1993,Bartana1997,Eitan2011,Schaefer2020}, as well as of many algorithms developed by Rabitz and coworkers~\cite{ZhuJCP98,ZhuJCP98b,Maday2003,Ohtsuki2004,OhtsukiPRA07,HoPRE10,LiaoPRA11}
There is continuously ongoing research and improvements on Krotov's method~\cite{Reich2012,Morzhin2018,Goerz2019}, 
being of particular interest the algorithms which provide accelerated convergence of quantum control~\cite{Maday2003,Reich2012} or strategies to smooth the control parameters of the driving field~\cite{Mennemann2015}. In this context, there are also studies on the landscape,~\ie, topological and geometric structure defined by the functional~\cite{RabitzScience04,Brif2010,MoorePRA11}.

The Quantum Optimal Control Theory (QOCT) 
is a well tested method that has proven to be efficient and accurate to control the dynamics of a quantum system~\cite{Tannor1992}. In this work we revisit the underlying theory to rigorously obtain the control variational equations. To do so, we also derive the Euler-Lagrange equations explaining in detail and pedagogically the theory of variations. The latter can be of interests for many physicist or chemist as well as grade students who do not fully understand the theory of variations and can benefit of it to, for instance, develop numerical methods.

This paper is structured as follows: In Section~\ref{sec:optimal_control}, we apply the Euler-Lagrange equations to obtain a set of QOCT equations, also known as equations of motion, which differ in a term with the equations obtained in the literature~\cite{Peirce1988, Tannor1992, Werschnik2007}. In Section~\ref{sec:conditions_solution}, we present the necessary conditions to obtain continuous solutions and demonstrate that the previously labeled "canonical solutions" are discontinuous in the costate, but not in the wavefucntion or the driving field under certain conditions. Our concluding remarks are summarized in Section~\ref{sec:conclusions}. Additionally, in the appendix, we provide a brief introduction to theory of variations, derivations of the Euler-Lagrange equations, and explain how to impose constraints using a Lagrange multipliers. We also include some proofs relevant to QOCT mentioned in the primary text.

\section{Quantum Optimal control}
\label{sec:optimal_control}

Quantum optimal control is a family of methods to optimize a specific observable for a quantum system. This involves imposing a cost, which is a constraint or condition on the field driving the system. In this section, we derive the QOCT
equations by utilizing the Euler-Lagrange equations (ELE). The ELE are derived using the theory of variations, and their solutions correspond to the stationary functions. Specifically, the ELE corresponding to the functional 
\begin{equation}
    \int_\Omega \mathcal{F}[y,y_{x_1},y_{x_2},\ldots,y_{x_n}, x_1,x_2,\ldots,x_n,\Omega^\prime]\mathrm{d}\Omega^\prime 
\end{equation}
are given by
\begin{equation}
    \label{eq:euler_lagrange_higher_order_in_text}
     \dfrac{\partial}{\partial y}\mathcal{F}+ \sum_{k=1}^m(-1)^k\sum\limits_{j=1}^n\dfrac{\mathrm{d^k}}{\mathrm{d}x^k_j}\dfrac{\partial}{\partial y_{x^k_j}}\mathcal{F}=0,
\end{equation}
being $y_{x_j^m}(x_1,x_2,\ldots,x_n)\equiv\frac{\partial^m}{\partial {x_j}^m}y$. The ELE are derived in \ref{sec:theory_of_variations} and \ref{sec:derivation_euler_lagrange_higher_order}.

\begin{figure}
    \centering
    \includegraphics[width=.75\linewidth]{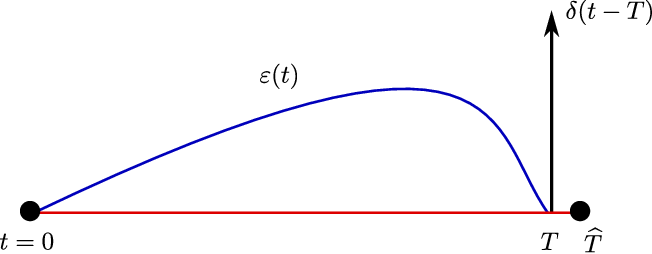}
    \caption{Sketch of a time interval illustrating the definition of $T\text{ and }\widehat{T}$.}
    \label{fig:time_interval}
\end{figure}

Our objective is to optimize the expectation value of the observable $O$ while minimizing a cost functional and satisfying the TDSE. To achieve this, we define a functional $J$ and then its extremal points. The functional $J$ may be divided into three terms:
\begin{eqnarray}
    \label{eq:jtotal}
&&J[\psi,\psi^*,\chi,\chi^*,\varepsilon]=J_\text{opt}[\psi,\psi^*]+J_\text{cost}[\varepsilon]+J_\text{TDSE}[\psi,\psi^*,\chi,\chi^*,\varepsilon].
\end{eqnarray}
First, the functional corresponding to optimizing the expectation value of the operator $O$
\begin{equation}
    \label{eq:optimize}
    J_\text{opt}[\psi,\psi^*]=\int_\Omega \psi^*(\Omega, T)O\psi(\Omega, T)\mathrm{d}\Omega,
\end{equation}
where $\psi(\Omega,t)$ is the wavefunction which describes the quantum system and $\Omega$ the degrees of freedom (Euler angles, cartesian coordinates of one or many-body,~\ldots). Then, we set the \emph{cost},~\ie, a condition on the field, as~\cite{Werschnik2007}
\begin{equation}
    \label{eq:cost}
    J_\text{cost}[\varepsilon]=-\alpha\int\limits_0^{T} \left[\varepsilon(t)-\varepsilon_\text{ref}(t)\right]^2\mathrm{d}t,
\end{equation}
 where $\alpha$ is a penalty factor. It is worth noting that $J_\text{cost}[\varepsilon]$ may be more complex by including a mask function~\cite{Werschnik2007}. However, for the sake of simplicity and without loss of generality, we will consider this simpler case. Finally, the functional which ensures that the TDSE is fulfilled reads as
\begin{equation}
    \label{eq:tdse}
    J_\text{TDSE}[\psi,\psi^*,\chi,\chi^*,\varepsilon]=-2\operatorname{Im}\int\limits_0^{\widehat{T}}\left[\int_\Omega\chi^*(\Omega,t)\left(i\frac{\partial}{\partial t}-H(\Omega, \varepsilon)\right)\psi(\Omega,t)\mathrm{d}\Omega\right]\mathrm{d}t,
\end{equation}
where $\chi(\Omega,t)$ is the Lagrange multiplier. In this work, we set the Hamiltonian to $H=-\frac{1}{2}\nabla^2+V(\Omega,\varepsilon(t))$, where $V(\Omega,\varepsilon(t))$ is the internal potential plus the interaction with a driving field, $\varepsilon(t)$. In contrast to the literature, we take the upper limit of the time interval to $\widehat{T}$, a value greater than the time at which the function to optimize is evaluated, $T$, denoted as the~\emph{measurement time}~\cite{Tannor1992}. Specifically, we choose $\widehat{T}>T$, as shown in Fig.\ref{fig:time_interval}, to effectively manage the discontinuity caused by the Dirac delta in the QOCT equations, as described below.

To obtain the extremal points we derive the corresponding set of Euler-Lagrange equations~\eqref{eq:euler_lagrange_higher_order_in_text}.
In the case of $\varepsilon(t)$, we find
    \begin{eqnarray}
     &&2\operatorname{Im}\int_\Omega\chi^*(\Omega,t)\frac{\partial}{\partial\varepsilon}H(\Omega, \varepsilon)\psi(\Omega,t)\mathrm{d}\Omega-2\alpha\left[(\varepsilon(t)-\varepsilon_\textup{ref}(t)\right]=0\Rightarrow\\
     \label{eq:field_qoc}
     && \varepsilon(t)=\varepsilon_\textup{ref}(t)+\frac{1}{\alpha}\operatorname{Im}\int_\Omega\chi^*(\Omega,t)\frac{\partial}{\partial\varepsilon}V(\vec{r}, \varepsilon)\psi(\Omega,t)\mathrm{d}\Omega.
    \end{eqnarray}
    Now, by taking variations on the wavefunction $\psi(\Omega,t)$, we determine the Lagrange multiplier $\chi^*(\Omega,t)$. Note that only the part of the functional~\eqref{eq:tdse} that has a direct dependence on $\psi(\Omega,t)$ is relevant. As $\psi(\Omega,t)$ and $\psi^*(\Omega,t)$ are independent of each other (for further information, refer to~\ref{sec:independence_of_conjugate_functions}), the relevant term reads as
    \begin{equation}
    j_\psi[\chi^*,\varepsilon,\psi,\Omega,t]=i\chi^*(\Omega,t)\left(i\frac{\partial}{\partial t}-H(\Omega,\varepsilon)\right)\psi(\Omega,t)+\psi^*(\Omega,t)O\psi(\Omega,t)\delta(t-T).
    \end{equation}
    Then, the terms in the ELE are
    \begin{eqnarray}
    \dfrac{\partial j_\psi}{\partial \psi}&=&i\chi^*(\Omega,t)\left[-V(\Omega,\varepsilon)\right]+O\psi^*(\Omega,t)\delta(t-T),\\
    \dfrac{\partial j_\psi}{\partial \dot\psi}&=&-\chi^*(\Omega,t),\quad  \dfrac{\mathrm{d}}{\mathrm{d}t}\dfrac{\partial j_\psi}{\partial \dot\psi}=-\dot\chi^*(\Omega,t),\\
    \dfrac{\partial j_\psi}{\partial \psi_{x_j}}&=&0,\quad \dfrac{\partial j_\psi}{\partial \psi_{x_j^2}}=\cfrac{i}{2}\chi^*(\Omega,t),\quad \dfrac{\mathrm{d}^2}{\mathrm{d}x_j^2}\dfrac{\partial j_\psi}{\partial \psi_{x_j^2}}=\cfrac{i}{2}\chi_{x_j^2}^*(\Omega,t)~\text{for all}~x_j,
    \end{eqnarray}
    where we use the notation $\dot{\psi}(\Omega,t)\equiv\frac{\partial}{\partial t}\psi(\Omega,t)$ and ${\psi}_{x_j^n}(\Omega,t)\equiv\frac{\partial^n}{\partial x_j^n}\psi(\Omega,t)$.
    Note that we have used the identity $\int_\Omega \psi^*(\Omega, T)O\psi(\Omega, T)\mathrm{d}\Omega=\int_\Omega \int_0^{\widehat{T}}\psi^*(\Omega, t)O\psi(\Omega, t)\delta(t-T)\mathrm{d}\Omega\mathrm{d}t$ and Eq.~\eqref{eq:euler_lagrange_higher_order_in_text} to derive the contributions to the ELE with high order derivatives. After summing all the terms, we obtain
    \begin{equation}
    \label{eq:tdse_chi_star}
        \left(i\dfrac{\partial}{\partial t}+H(\Omega,\varepsilon)\right)
\chi^*(\Omega,t)=-iO\psi^*(\Omega,t)\delta(t-T).
    \end{equation}
        To obtain the ELE for $\chi(\Omega,t)$ we can either take variations on $\psi^*(\Omega,t)$ or simply take the complex conjugate of Eq.~\eqref{eq:tdse_chi_star}, 
    \begin{equation}
    \label{eq:tdse_chi}
        \left(i\dfrac{\partial}{\partial t}-H(\Omega,\varepsilon)\right)
\chi(\Omega,t)=-iO\psi(\Omega,t)\delta(t-T).
    \end{equation}
The TDSE of $\psi^*(\Omega,t)$ is obtained by taking variations of $\chi(\Omega,t)$.
The relevant term reads 
    \begin{equation}
    j_{\chi}[\psi^*,\chi,\varepsilon,\Omega,t]=-i
    \chi(\Omega,t)\left(i\frac{\partial}{\partial t}+H(\Omega, \varepsilon)\right)\psi^*(\Omega,t)
    \end{equation}
    Each term in the ELE is
    \begin{eqnarray}
    \dfrac{\partial j_{\chi}}{\partial\dot\chi}&=&0,\qquad \dfrac{\mathrm{d}}{\mathrm{d}t}\dfrac{\partial j_{\chi}}{\partial\dot\chi}=0\\
    \dfrac{\partial j_{\chi}}{\partial\chi}&=&-i\left(i\frac{\partial}{\partial t}+H(\Omega,\varepsilon(t))\right)\psi^*(\Omega,t).
    \end{eqnarray}
    By combining all the terms we derive the EOM
    \begin{equation}
    \label{eq:tdse_psi_star}
        \left(i\frac{\partial}{\partial t}+H(\Omega, \varepsilon)\right)\psi^*(\Omega,t)=0
    \end{equation}
        Finally, we take variations with respect to the Lagrange multiplier $\chi^*(\Omega,t)$. As done above, we work with the relevant term 
     \begin{eqnarray}
    j_{\chi^*}[\psi,\chi^*,\varepsilon,\Omega,t]
    &=& i\chi^*(\Omega,t)\left(i\frac{\partial}{\partial t}-H(\Omega, \varepsilon)\right)\psi(\Omega,t).
    \end{eqnarray}
    The terms in the associated ELE are
    \begin{eqnarray}
    \dfrac{\partial j_{\chi}}{\partial\dot\chi^*}&=&0,\qquad \dfrac{\mathrm{d}}{\mathrm{d}t}\dfrac{\partial j_{\chi^*}}{\partial\dot\chi^*}=0,\\
    \dfrac{\partial j_{\chi^*}}{\partial\chi^*}&=&i\left(i\frac{\partial}{\partial t}-H(\Omega, \varepsilon)\right)\psi(\Omega,t) \ ,
    \end{eqnarray}
    from which we obtain the TDSE for $\psi(\Omega,t)$
    \begin{equation}
    \label{eq:tdse_psi}
        \left(i\frac{\partial}{\partial t}-H(\Omega, \varepsilon)\right)\psi(\Omega,t)=0.
    \end{equation}
    As expected, Eq.~\eqref{eq:tdse_psi} is the complex conjugate of Eq.~\eqref{eq:tdse_psi_star}, which ensures that the functional $J[\psi,\psi^*,\chi,\chi^*,\varepsilon]$ is appropriate.
Summing up, we derived the set of equations
\begin{eqnarray}
\label{eq:psi_eom}
&&\left(i\dfrac{\partial}{\partial t}-H(\Omega,\varepsilon)\right)
\psi(\Omega,t)=0,\\
\label{eq:chi_eom}
&&\left(i\dfrac{\partial}{\partial t}-H(\Omega,\varepsilon)\right)
\chi(\Omega,t)=-iO\psi(\Omega,t)\delta(t-T),\\
\label{eq:epsilon_eom}
 &&\varepsilon(t)=\varepsilon_\textup{ref}(t)+\frac{1}{\alpha}\operatorname{Im}\int_\Omega\chi^*(\Omega,t)\frac{\partial}{\partial\varepsilon}H(\Omega, \varepsilon)\psi(\Omega,t)\mathrm{d}\Omega,
\end{eqnarray}
and their complex conjugates. 

In the following sections we obtain two different types of solutions, namely, continuous and discontinuous solutions. The latter are related to the \emph{canonical} QOCT equations.

\section{Extremal trajectories: Conditions and continuity}
\label{sec:conditions_solution}

\subsection{Continuous solutions}
\label{sec:continuous_solution}

In Eq.\eqref{eq:psi_eom}, the continuity of $\psi(\Omega,t)$ is guaranteed, while the continuity of $\chi(\Omega,t)$ and thus $\varepsilon(t)$ cannot be assured by Eq.\eqref{eq:chi_eom}. Here, we show that $\chi(\Omega,t)$ is continuous if it fulfills $\chi(\Omega,t)=\frac{i}{2\pi n}O\psi(\Omega,t)$, where $n\in\mathds{Z}-{0}$.

\begin{proof}
Let us assume the ansatz $\chi(\Omega,t)=\widetilde{\chi}(\Omega,t)e^{i\phi\Theta(t-T)}$, where $\widetilde{\chi}(\Omega,t)$ is continuous and fullfils the TDSE, $\phi\in \mathds{C}$ and $\Theta(t-T)$ is the Heaviside function. We substitute the ansatz in Eq.~\eqref{eq:chi_eom}
\begin{align*}
&i\frac{\partial}{\partial t}\left(\widetilde{\chi}(\Omega,t)e^{i\phi\Theta(t-T)}\right)-\hat{H}(\Omega,t)\widetilde{\chi}(\Omega,t)e^{i\phi\Theta(t-T)}=-iO\psi(\Omega,t)\delta(t-T)\\
&ie^{i\phi\Theta(t-T)}\frac{\partial}{\partial t}\widetilde{\chi}(\Omega,t)+i\widetilde{\chi}(\Omega,t)\frac{\partial}{\partial t}e^{i\phi\Theta(t-T)}-\hat{H}(\Omega,t)\widetilde{\chi}(\Omega,t)e^{i\phi\Theta(t-T)}=-iO\psi(\Omega,t)\delta(t-T)\\
&e^{i\phi\Theta(t-T)}\left(i\frac{\partial}{\partial t}\widetilde{\chi}(\Omega,t)-\hat{H}(\Omega,t)\widetilde{\chi}(\Omega,t)\right)=\left(\phi\widetilde{\chi}(\Omega,t)e^{i\phi\Theta(t-T)}-iO\psi(\Omega,t)\right)\delta(t-T)
\end{align*}
To remove the discontinuity, we may set 
\begin{equation}
    \lim\limits_{t\rightarrow T}\phi\widetilde{\chi}(\Omega,t)e^{i\phi\Theta(t-T)}=iO\psi(\Omega,T)\Rightarrow \lim\limits_{t\rightarrow T}\chi(\Omega,t)=\frac{i}{\phi}O\psi(\Omega,T)
\end{equation}
Finally, $\lim\limits_{t\rightarrow T}\widetilde{\chi}(\Omega,t)e^{i\phi\Theta(t-T)}$ is defined if $e^{i\phi\Theta(t-T)}$ is continuous,~\ie, $\phi=2\pi n$ with $n\in\mathds{Z}-\{0\}$. Therefore, $\chi(\Omega,t)$ is continuous if and only if $\chi(\Omega,T)=\frac{i}{2\pi n}O\psi(\Omega,T)$. Note that it is easy to prove that $\epsilon(t)$ is also continuous since $\psi(\Omega,t)$ and $\chi(\Omega,t)$ are continuous.
\end{proof}

It should be noted that the condition of continuity is imposed at a specific time $T$, which poses a challenge as $T$ is not a boundary of the time interval (as illustrated in Fig.~\ref{fig:sketch_solution}). Therefore, such solutions may not be appropriate from a formal standpoint, especially when extending the system to a finite \emph{measurement time}. Let us also remark that the solution here obtained differs from the costate provided in the literature~\cite{Tannor1992, Werschnik2007, Schaefer2020}. In the following section, we will provide evidence to demonstrate that the costate reported in earlier studies corresponds to a solution that is discontinuous.

\subsection{Relationship to \emph{Canonical} quantum optimal control equations}
\label{sec:relationship_qoc_canonical}
In this section, we introduce a less restrictive condition for $\chi(\Omega,t)$ at $t=T$, allowing for discontinuity, and impose an initial condition such that $\chi(\Omega,T^\prime)=0$ for $T^\prime \textgreater{T}$.
\begin{proof}

Let us define $\chi(\Omega,t)=\widetilde{\chi}(\Omega,t)f(t)$, where $\widetilde{\chi}(\Omega,t)$ satisfies the TDSE $\left(i\dfrac{\partial}{\partial t}-H(\Omega,\varepsilon)\right)\widetilde{\chi}(\Omega,t)=0$ and $\widetilde{\chi}(\Omega,t)=\chi(\Omega,t)$ for $t\in [0,T)$. We plug it in Eq.~\eqref{eq:tdse_chi}
  \begin{eqnarray}
    \nonumber
        &&\left(i\dfrac{\partial}{\partial t}-H(\Omega,\varepsilon)\right)
\left(\widetilde{\chi}(\Omega,t)f(t)\right)=-iO\psi(\Omega,t)\delta(t-T)\\
\nonumber
&&i\widetilde{\chi}(\Omega,t)\dfrac{\partial}{\partial t}f(t)+f(t)\left(i\dfrac{\partial}{\partial t}\widetilde{\chi}(\Omega,t)-H(\Omega,\varepsilon)
\widetilde{\chi}(\Omega,t)\right)=-iO\psi(\Omega,t)\delta(t-T)\\
    \label{eq:tdse_chi_krotov}
    &&\widetilde{\chi}(\Omega,t)\dfrac{\partial}{\partial t}f(t)=-O\psi(\Omega,t)\delta(t-T)
    \end{eqnarray}
Next, we integrate over time in the interval $[T-\delta,T+\delta]$ with $\delta>0$ and take the limit $\delta\rightarrow 0$
\begin{equation}
    \label{eq:tdse_chi_krotov_delta}
    \lim_{\delta\rightarrow 0}f(\Omega,T+\delta)-\lim_{\delta\rightarrow 0}f(\Omega,T-\delta)=-\cfrac{O\psi(\Omega,T)}{\widetilde{\chi}(\Omega,T)}.
\end{equation}
Setting $\chi(\Omega,t)=0$ for $t>T$ and using that $\chi(\Omega,t)=\widetilde{\chi}(\Omega,t)$ for $t<T$, that is to say, $\lim_{\delta\rightarrow 0}f(T+\delta)=0$ and $\lim_{\delta\rightarrow 0}f(T-\delta)=1$, we obtain
\begin{equation}
    -1=-\cfrac{O\psi(\Omega,T)}{\widetilde{\chi}(\Omega,T)}\Rightarrow \widetilde{\chi}(\Omega,T)=O\psi(\Omega,T),
\end{equation}
therefore, we obtain the boundary condition for the costate
\begin{equation}
    \label{eq:tdse_chi_krotov_tminus}
    \lim_{\delta\rightarrow 0}\chi(\Omega,T-\delta)=O\psi(\Omega,T).
\end{equation}
Next, we plug Eq.~\eqref{eq:tdse_chi_krotov_tminus} in Eq.~\eqref{eq:epsilon_eom}
\begin{eqnarray}
\nonumber
\lim_{\delta\rightarrow 0}\varepsilon(T-\delta)&=&
\varepsilon_\textup{ref}(T)+\frac{1}{\alpha}\operatorname{Im}\int_\Omega O\psi^*(\Omega,T)\frac{\partial}{\partial\varepsilon}H(\Omega, \varepsilon)\psi(\Omega,T)\mathrm{d}\Omega\\
\label{eqn:varepsilon_tminus}
&=& \varepsilon_\textup{ref}(T)+\frac{1}{\alpha}\operatorname{Im}\melement{\psi}{O\frac{\partial}{\partial\varepsilon}H(\Omega, \varepsilon)}{\psi}
\end{eqnarray}
Note that the integral in Eq.~\eqref{eqn:varepsilon_tminus} is real if $O\frac{\partial}{\partial\varepsilon}H(\Omega, \varepsilon)$ is Hermitian, that is, $\left[O,\frac{\partial}{\partial\varepsilon}H(\Omega, \varepsilon)\right]=0$. As a consequence, we have $\lim\limits_{\delta\rightarrow 0}\varepsilon(T-\delta)=\varepsilon_\text{ref}(T)$. Moreover, even though $\chi(\Omega,t)$ is not continuous at $t=T$, we have that $\lim_{\delta\rightarrow 0}\varepsilon(T+\delta)=\varepsilon_\text{ref}(T)$, which ensures the continuity of $\varepsilon(t)$ at $t=T$.
\end{proof}

\section{Conclusions}
\label{sec:conclusions}
 In this study, we have presented a comprehensive explanation of the theory of variations that is essential to derive the Quantum Optimal Control Theory (QOCT) equations. Specifically, we have derived the Euler-Lagrange equations for multiple variables and demonstrated a systematic approach to incorporate constraints in any multidimensional functional by utilizing the Lagrange multipliers.

Moreover, we have shown that the \emph{canonical} solutions of the QOCT equations, which are commonly reported in the literature, exhibit discontinuity at the measurement time, as it is commonly assumed. In contrast, we have identified a new set of continuous solutions of QOCT equations. However, the lack of well-defined boundary conditions limits their practical applicability.

Finally, let us highlight that our work presents a foundation for developing novel optimal control strategies using our methodology.

\ack
J.J.O. gratefully acknowledges the funding from the Madrid Government (Comunidad de Madrid Spain) under the Multiannual Agreement with Universidad Complutense de Madrid in the line Research Incentive for Young PhDs, in the context of the V PRICIT (Regional Programme of Research and Technological Innovation) (Grant: PR27/21-010), Project PID2019-106732GB-I00 (MINECO) and the Social European Fund and Juan de la Cierva-Incorporaci\'on granted by the Ministerio de Ciencia e Innovaci\'on. I.R.S. acknowledges support from MINECO PID2021-122796NB-I00. K.C. acknowledges support form MINECO PID2020-118180GB-I00, and Junta de Andalucía (Spain) grants A-FQM-441-UGR18 and P20-00164.

\appendix

\section{Theory of variations}
\label{sec:theory_of_variations}

As we have sketched above, the QOCT is a widely used procedure to maximize or minimize an observable of a quantum system by setting several constraints. 
Most of the theory developed in the literature addresses physicists and chemists, and lacks a rigorous mathematical framework, which demands clear and correct proofs. In particular, this is due to a misunderstanding in the concept of~\emph{variation}, which is defined in mathematics as~\emph{any function $\eta(\Omega)$ that perturbs a given function $f(\Omega)$, fulfilling that $\eta(\Omega)$ vanishes in the border of the domain of $f(\Omega)$}. Let us also note that this vanishing condition may also apply for the derivatives of $\eta(\Omega)$, as we explain in detail in~\ref{sec:derivation_euler_lagrange_higher_order}. 

In the following sections we show and derive the mathematical tools to correctly derive the EOM of QOC. We first derive the Euler-Lagrange equations (ELE), which are the equations fulfilled by the functions which optimize a given functional. By applying the ELE, we do not need to make explicit use of the variations, facilitating and clarifying the derivation of our set of QOCT equations. Next, we explain in detail how to add constraints to the functional, such as the total energy of the driving field. 

\subsection{Derivation of Euler-Lagrange equations}
\label{sec:derivation_euler_lagrange}

 The minimization principle consists on seeking for the minimum of the functional
 \begin{equation}
\label{eq:functional}
\int_\Omega \mathcal{F}[\widehat{y},\widehat{y}_{x_1},\widehat{y}_{x_2},\ldots,\widehat{y}_{x_n},\Omega^\prime]\mathrm{d}\Omega^\prime,
\end{equation}
where $\mathcal{F}$ is a functional of $\widehat{y}(x_1,x_2,\ldots,x_n)$ defined in a $n$ dimensional domain $\Omega$, its first derivatives and $\Omega^\prime=(x_1,x_2,\ldots,x_n)$. Note that we use the notation $y_{x_i}=\frac{\partial y}{\partial x_{i}}$. 
The Euler-Lagrange equations (ELE) are the conditions of the extremal functions, that is, the conditions that have to be satisfied by $\{y_{x_j}\}_{j=1}^n$, that optimize the functional~\eqref{eq:functional}. They read as
\begin{eqnarray}
\nonumber
    &&\dfrac{\partial}{\partial \widehat{y}}\mathcal{F}(\widehat{y},\widehat{y}_{x_1},\ldots,\widehat{y}_{x_n},x_1,\ldots,x_n)- \\
        \label{eq:euler_lagrange_general_before_proof}
     &&   \sum\limits_{j=1}^n\dfrac{\mathrm{d}}{\mathrm{d}x_j}\dfrac{\partial}{\partial \widehat{y}_{x_j}}\mathcal{F}(\widehat{y},\widehat{y}_{x_1},\ldots,\widehat{y}_{x_n},x_1,\ldots,x_n)=0.
\end{eqnarray}
\begin{proof}
To seek for these extremal functions we follow the same strategy than Lagrange~\cite{Goldstine1980}. We take 
\begin{equation}
\label{eq:y_plus_variation}
    \widehat{y}(x_1,x_2,\ldots,x_n,\epsilon)=y(x_1,x_2,\ldots,x_n)+\xi\eta(x_1,x_2,\ldots,x_n),
\end{equation}
where $\eta(x_1,x_2,\ldots,x_n)\in \mathds{C}$ is such that vanishes on  $\partial\Omega$,~\ie, the border of the domain $\Omega$ and $\epsilon\in \mathds{R}$. Let us note that we may find in the literature that the~\emph{variation} is defined as $\delta y(x_1,x_2,\ldots,x_n)\equiv\xi\eta(x_1,x_2,\ldots,x_n)$. Thus, $\widehat{y}(x_1,x_2,\ldots,x_n,\epsilon)$ and $y(x_1,x_2,\ldots,x_n)$ have the same value in the $\partial\Omega$ for all $\eta(x_1,x_2,\ldots,x_n)$. In Fig.~\ref{fig:sketch_solution}, we illustrate how the variations affect a curve in a plane with the calculation of the shortest distance between two points. For the sake of simplicity we restrict our analysis to the 2 dimensional case, which can be easily generalized. The stationary function is a geodesic which corresponds to the straight line linking two points. Therefore, the problem is reduced to finding $y(x_1,x_2,\ldots,x_n)$ so that
\begin{equation}
\label{eq:functional_partial_epsilon}
    \left.\dfrac{\partial}{\partial \xi}\int_\Omega\mathcal{F}[\widehat{y},\widehat{y}_{x_1},\widehat{y}_{x_2},\Omega^\prime]\mathrm{d}\Omega^\prime\right|_{\xi=0}=0.
\end{equation}

\begin{figure}
\centering
\includegraphics[width=.6\linewidth]{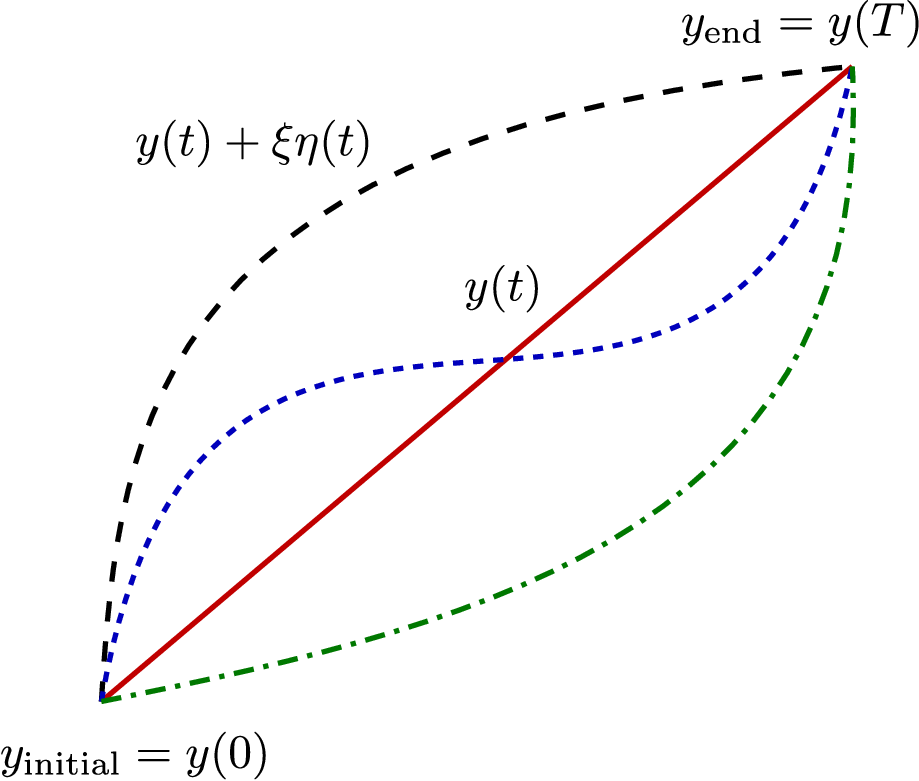}
\caption{\label{fig:sketch_solution} Paths linking $y_\text{initial}$ and $y_\text{end}$. The function $y(t)$ (solid red line) represents the extremal function and the other curves correspond to this extremal solution plus a variation.}
\end{figure}

We assume that $y(x_1,x_2)$ is a stationary point of $\int_\Omega\mathcal{F}[y,y_{x_1},y_{x_2},x_1,x_2]\mathrm{d}x_1\mathrm{d}x_2$. Then, we study the functional close to this point by adding a~\emph{variation} $\delta y=\xi\eta(x_1,x_2)$, where $\eta(x_1,x_2)$ is any function which vanishes at the border of the domain of $y(x_1,x_2)$ (see Fig.~\ref{fig:sketch_solution}). The variation ensures that for $\xi=0$, $\int_\Omega\mathcal{F}[y+\xi\eta,y_{x_1}+\xi\eta_{x_1},y_{x_2}+\xi\eta_{x_2},x_1,x_2]\mathrm{d}x_1\mathrm{d}x_2$ has an extremal,~\ie, it fulfills the condition
\begin{equation}
\label{eq:functional_partial_epsilon_12}
    \left.\dfrac{\partial}{\partial \xi}\int_\Omega\mathcal{F}[\widehat{y},\widehat{y}_{x_1},\widehat{y}_{x_2},x_1,x_2]\mathrm{d}x_1\mathrm{d}x_2\right|_{\xi=0}=0.
\end{equation}
 Expanding the derivative, we obtain
\begin{eqnarray}
&&\int\limits_\Omega\left[\dfrac{\partial}{\partial \widehat{y}}\mathcal{F}(\widehat{y},\widehat{y}_{x_1},\widehat{y}_{x_2},x_1,x_2)\dfrac{\partial \widehat{y}}{\partial \xi}+\dfrac{\partial}{\partial \widehat{y}_{x_1}}\mathcal{F}(\widehat{y},\widehat{y}_{x_1},\widehat{y}_{x_2},x_1,x_2)\dfrac{\partial \widehat{y}_{x_1}}{\partial \xi}+\right.\\
&&\left.\dfrac{\partial}{\partial \widehat{y}_{x_2}}\mathcal{F}(\widehat{y},\widehat{y}_{x_1},\widehat{y}_{x_2},x_1,x_2)\dfrac{\partial \widehat{y}_{x_2}}{\partial \xi}\right]\mathrm{d}x_1\mathrm{d}x_2=\\
&&\int\limits_\Omega\left[\dfrac{\partial}{\partial \widehat{y}}\mathcal{F}(\widehat{y},\widehat{y}_{x_1},\widehat{y}_{x_2},x_1,x_2)\eta(x_1,x_2)+\right.\\
&&\left.\dfrac{\partial}{\partial \widehat{y}_{x_1}}\mathcal{F}(\widehat{y},\widehat{y}_{x_1},\widehat{y}_{x_2},x_1,x_2)\eta_{x_1}(x_1,x_2)+\right.\\
&& \left.\dfrac{\partial}{\partial \widehat{y}_{x_2}}\mathcal{F}(\widehat{y},\widehat{y}_{x_1},\widehat{y}_{x_2},x_1,x_2)\eta_{x_2}(x_1,x_2)\right]\mathrm{d}x_1\mathrm{d}x_2=\\
&& \int\limits_\Omega\dfrac{\partial}{\partial \widehat{y}}\mathcal{F}(\widehat{y},\widehat{y}_{x_1},\widehat{y}_{x_2},x_1,x_2)\eta(x_1,x_2)\mathrm{d}x_1\mathrm{d}x_2+\\
\label{eq:functional_depsilon_extend_x1}
&& \left[\dfrac{\partial}{\partial \widehat{y}_{x_1}}\mathcal{F}(\widehat{y},\widehat{y}_{x_1},\widehat{y}_{x_2},x_1,x_2)\eta(x_1,x_2)\right]_{\partial \Omega}\\
&&-\int\limits_\Omega\mathrm{d}x_1\mathrm{d}x_2\dfrac{\mathrm{d}}{\mathrm{d}x_1}\dfrac{\partial}{\partial \widehat{y}_{x_1}}\mathcal{F}(\widehat{y},\widehat{y}_{x_1},\widehat{y}_{x_2},x_1,x_2)\eta(x_1,x_2)+\\
\label{eq:functional_depsilon_extend_x2}
&& \left[\dfrac{\partial}{\partial \widehat{y}_{x_2}}\mathcal{F}(\widehat{y},\widehat{y}_{x_1},\widehat{y}_{x_2},x_1,x_2)\eta(x_1,x_2)\right]_{\partial \Omega} \\
&& -\int\limits_\Omega\dfrac{\mathrm{d}}{\mathrm{d}x_2}\dfrac{\partial}{\partial \widehat{y}_{x_2}}\mathcal{F}(\widehat{y},\widehat{y}_{x_1},\widehat{y}_{x_2},x_1,x_2)\eta(x_1,x_2)\mathrm{d}x_1\mathrm{d}x_2.
\end{eqnarray}
 The terms~\eqref{eq:functional_depsilon_extend_x1} and~\eqref{eq:functional_depsilon_extend_x2} vanish because $\eta(x_1,x_2)=0$ in the border of the domain,~\ie, for $(x_1,x_2)\in \partial\Omega$. Therefore, the condition for the stationary point is
\begin{eqnarray}
\nonumber
   && \int\limits_\Omega\left[ \dfrac{\partial}{\partial \widehat{y}}\mathcal{F}(\widehat{y},\widehat{y}_{x_1},\widehat{y}_{x_2},x_1,x_2)-  \dfrac{\mathrm{d}}{\mathrm{d}x_1}\dfrac{\partial}{\partial \widehat{y}_{x_1}}\mathcal{F}(\widehat{y},\widehat{y}_{x_1},\widehat{y}_{x_2},x_1,x_2) \right.\\
       \label{eq:functional_condition}
&&   \left.-\dfrac{\mathrm{d}}{\mathrm{d}x_2}\dfrac{\partial}{\partial \widehat{y}_{x_2}}\mathcal{F}(\widehat{y},\widehat{y}_{x_1},\widehat{y}_{x_2},x_1,x_2)\right]\eta(x_1,x_2)\mathrm{d}x_1\mathrm{d}x_2=0.
\end{eqnarray}
Since this condition has to be fulfilled for all $\eta(x_1,x_2)$
\begin{equation}
    \label{eq:euler_lagrange}
    \dfrac{\partial}{\partial \widehat{y}}\mathcal{F}(\widehat{y},\widehat{y}_{x_1},\widehat{y}_{x_2},x_1,x_2)-  \dfrac{\mathrm{d}}{\mathrm{d}x_1}\dfrac{\partial}{\partial \widehat{y}_{x_1}}\mathcal{F}(\widehat{y},\widehat{y}_{x_1},\widehat{y}_{x_2},x_1,x_2)-  \dfrac{\mathrm{d}}{\mathrm{d}x_2}\dfrac{\partial}{\partial \widehat{y}_{x_2}}\mathcal{F}(\widehat{y},\widehat{y}_{x_1},\widehat{y}_{x_2},x_1,x_2)=0.
\end{equation}
This result can be generalized for more than two variables following the same procedure, obtaining
\begin{equation}
    \label{eq:euler_lagrange_general}
    \dfrac{\partial}{\partial y}\mathcal{F}(y,y_{x_1},\ldots,y_{x_n},x_1,\ldots,x_n)-  \sum\limits_{j=1}^n\dfrac{\mathrm{d}}{\mathrm{d}x_j}\dfrac{\partial}{\partial y_{x_j}}\mathcal{F}(y,y_{x_1},\ldots,y_{x_n},x_1,\ldots,x_n)=0,
\end{equation}
where we have used that $\widehat{y}(\Omega,t)=y(\Omega,t)$ for $\xi=0$.
\end{proof}
It is important to remark that it is not necessary to know  $y$ in the entire border if the ELE guarantee the existence and uniqueness of the solution. 

Finally, let us now consider a functional depending on a function of high order derivatives as $\mathcal{F}(\widehat{y},\{\widehat{y}_{x_j}\}_{j=1}^n,\{\widehat{y}_{x_j^2}\}_{j=1}^n,\ldots,\{\widehat{y}_{x_j^m}\}_{j=1}^n,x_1,\ldots,x_n)$, where $y_{x_j^k}\equiv\cfrac{\partial ^ky}{\partial x^k_j}$. Thus, the ELE for functional with higher order derivatives read
\begin{equation}
    \label{eq:euler_lagrange_higher_order}
     \dfrac{\partial}{\partial y}\mathcal{F}+ \sum_{k=1}^m(-1)^k\sum\limits_{j=1}^n\dfrac{\mathrm{d^k}}{\mathrm{d}x^k_j}\dfrac{\partial}{\partial y_{x^k_j}}\mathcal{F}=0,
\end{equation}
where we have not explicitly indicated the arguments of the functional for the sake of clarity. A detailed proof can be found in~\ref{sec:derivation_euler_lagrange_higher_order}.

\subsection{Adding constraints}
\label{sec:adding_constraints}

Adding a constraint may be mandatory to solve a given problem, and it is done by using the Lagrange multipliers~\cite{Stewart2011}. In the context of calculus, to find the stationary point of a function $y(x_1,x_2)$ fulfilling the condition $g(x_1,x_2)=0$ we have to find the stationary point of
\begin{equation}
\label{eq:lagrange_multipliers_calculus}
f(x_1,x_2,\lambda)=y(x_1,x_2)+\lambda g(x_1,x_2).
\end{equation}
The stationary points fulfill
\begin{equation}
    \dfrac{\partial}{\partial x_1}f(x_1,x_2,\lambda)=\dfrac{\partial}{\partial x_2}f(x_1,x_2,\lambda)=\dfrac{\partial}{\partial \lambda}f(x_1,x_2,\lambda)=0,
\end{equation}
where $\lambda$ is the~\emph{Lagrange} multiplier and the last equality ensures that the constraint $g(x_1,x_2)=0$. 

The extension to the theory of variations consists on taking a function as Lagrange multiplier. In this case, we restrict to functionals with first derivatives only. The generalization can be done following the steps described in~\ref{sec:derivation_euler_lagrange_higher_order}. The multiplier function, $\lambda(x_1,x_2,\ldots,x_n)$, may depend on an arbitrary number of variables, depending on the constraint. Let us show how to construct the functionals based on these constrains with some illustrative examples in a two dimensional space.

\subsubsection{Constraint at each $(x_1,x_2)\in\Omega$}

 First, we consider a constraint at each $(x_1,x_2)\in\Omega$,~\ie, $g(y,y_{x_1},y_{x_2},\ldots,x_1,x_2)=0$. Then, we sum the term 
    \begin{equation}
        \label{eq:condition_everypoint}
        \int\limits_\Omega \lambda g(y,y_{x_1},y_{x_2},x_1,x_2)\mathrm{d}\Omega.
    \end{equation}
    to the functional to be minimized, $\int\limits_\Omega \mathcal{F}(y,y_{x_1},y_{x_2},x_1,x_2)\mathrm{d}\Omega$.
    Now, we apply the ELE~\eqref{eq:euler_lagrange} to $\lambda(x_1,x_2)$ obtaining $g(y,y_{x_1},y_{x_2},\ldots,x_1,x_2)=0$. Note that we are not taking into account  $\int\limits_\Omega \mathcal{F}(y,y_{x_1},y_{x_2},\ldots,x_1,x_2)\mathrm{d}\Omega$ here, since it does not depend on $\lambda$. Finally, in order to obtain the ELE corresponding to $y$'s we use the functional with constrains
    \begin{eqnarray}
        \nonumber
        &&\mathcal{G}(y,y_{x_1},y_{x_2},x_1,x_2,\lambda,\epsilon,\eta)=\\
        &&\int\limits_\Omega \left[\mathcal{F}(y,y_{x_1},y_{x_2},x_1,x_2)+\lambda g(y,y_{x_1},y_{x_2},x_1,x_2)\right]\mathrm{d}\Omega
    \end{eqnarray}
    
    \subsubsection{Constraint given by an integral in one variable of a function of two variables}
    
    Now we analyze another remarkable kind of constrains, given by \emph{an integral in one variable of a function of two variables},~\ie, $\int_{\Omega_1}g(y,\{y_{x_i}\},\{y_{x_i,x_j}\},\ldots,x_1,x_2)\mathrm{d}x_1=0$. For example, the conservation of the norm of the wavefunction (integration in the position space of the norm squared) for all time $t$ is commonly applied to many-body physics such as Multiconfigurational Time-Dependent Hartree-Fock and related self-consistent methods~\cite{Beck2000a,Sato2013,Miyagi2013}.
    
    In this case we include the functional 
    \begin{equation}
        \label{eq:condition_int_onevariable}
        \int\limits_{\Omega_1}\int\limits_{\Omega_2}\lambda(x_2)g(y,y_{x_1},y_{x_1},x_1,x_2)\mathrm{d}x_1\mathrm{d}x_2,
    \end{equation}
    where we assume that $\Omega$ can be split in $\Omega_1$ and $\Omega_2$, with $x_j\in\Omega_j$.
    Here we can not apply directly the ELE as it is expressed above, since the multiplier depends on less variables than the dimensions of $\Omega$. However, we can still use Lagrange's methodology by adding a variation $\xi\eta(x_2)$ to the multiplier and deriving with respect to $\xi$. 
        \begin{eqnarray}
        \nonumber
        \mathcal{G}(y,y_{x_1},y_{x_2},x_1,x_2,\xi,\eta)&=&\int\limits_{\Omega_2}\int\limits_{\Omega_1}(\lambda(x_2)+\xi\eta(x_2))g(y,y_{x_1},y_{x_2},x_1,x_2)\mathrm{d}x_1\mathrm{d}x_2,
    \end{eqnarray}
    and then, deriving with respect to $\xi$
    \begin{eqnarray}
        \dfrac{\partial}{\partial \xi}\mathcal{G}(y,y_{x_1},y_{x_2},x_1,x_2,\xi,\eta)&=&\int\limits_{\Omega_2}\int\limits_{\Omega_1}\eta(x_2)g(y,y_{x_1},y_{x_2},x_1,x_2)\mathrm{d}x_1\mathrm{d}x_2\\
        \label{eq:condition_constraint_onevariable}
        &=&\int\limits_{\Omega_2}\eta(x_2)\int\limits_{\Omega_1}g(y,y_{x_1},y_{x_2},x_1,x_2)\mathrm{d}x_1\mathrm{d}x_2.
    \end{eqnarray}
Thus, $\dfrac{\partial}{\partial \xi}\mathcal{G}(y,y_{x_1},y_{x_2},x_1,x_2,\xi,\eta)=0$ for all $\eta(x_2)$ if and only if the constrain $\int\limits_{\Omega_1}g(y,y_{x_1},y_{x_2},x_1,x_2)\mathrm{d}x_1=0$ is fulfilled. 

\subsubsection{General case: Constraint given by an integral in $m$ variables of a function of $n$ variables} 

The strategy consists on including a Lagrange multiplier in the $n-m$ variables which are not integrated, as can be generalized from the previous cases. 

\section{Derivation of Euler-Lagrange type equation for higher order derivatives}
\label{sec:derivation_euler_lagrange_higher_order}

In this section we derive the ELE for a functional with higher order derivatives. First, we will derive the case of second and first order derivatives and one variable. Then, we will generalize the result.
\begin{proof}
Let us take the functional 
$\int\limits_\Omega \mathcal{F}(y,y_{x},y_{x^2},x)\mathrm{d}x$. We follow a similar procedure as in~\ref{sec:derivation_euler_lagrange} to compute the ELE. First, we evaluate the functional with the function $\widehat{y}(x)=y(x)+\xi\eta(x)$, where $y(x)$ is the extremal, $\eta(x)$ is a function such that $\eta(x)|_{\partial \Omega}=\eta_x(x)|_{\partial \Omega}=0$ and $\xi\in \mathds{R}$
\begin{eqnarray}
    \nonumber
    &&\mathcal{I}[(y+\xi\eta,y_{x}+\xi\eta_{x},y_{x^2}+\xi\eta_{x^2})]=\\
    &&\int\limits_\Omega \mathcal{F}(y+\xi\eta,y_{x}+\xi\eta_{x},y_{x^2}+\xi\eta_{x^2},x)\mathrm{d}x.
\end{eqnarray}
Since $y(x)$ is an extremal of the functional $\mathcal{I}$, then $\frac{\mathrm{d}}{\mathrm{d}\xi}\mathcal{I}[(y+\xi\eta,y_{x}+\xi\eta_{x},y_{x^2}+\xi\eta_{x^2})]|_{\xi=0}=0$. The derivative reads as
\begin{eqnarray}
\nonumber
&&\dfrac{\mathrm{d}}{\mathrm{d}\xi}\mathcal{I}[(y+\xi\eta,y_{x}+\xi\eta_{x},y_{x^2}+\xi\eta_{x^2})]=\\
\nonumber
&&\int\limits_\Omega\left[\dfrac{\partial}{\partial \widehat{y}}\mathcal{F}(\widehat{y},\widehat{y}_{x},\widehat{y}_{x}^2,x)\eta(x)+\dfrac{\partial}{\partial \widehat{y}_x}\mathcal{F}(\widehat{y},\widehat{y}_{x},\widehat{y}_{x}^2,x)\eta_x(x)\right.\\
\label{eq:functional_higher_derivatives_derivate_epsilon}
&&\dfrac{\partial}{\partial \widehat{y}_x^2}\mathcal{F}(\widehat{y},\widehat{y}_{x},\widehat{y}_{x^2},x)\eta_{x^2}(x)\left.\right]\mathrm{d}x,
\end{eqnarray}
where we have used that $\widehat{y}(x)=y(x)+\xi\eta(x)$ and therefore $\frac{\mathrm{d}}{\mathrm{d}\xi}\widehat{y}(x)=\eta(x)$. Now, we integrate by parts to remove any derivative of $\eta(x)$
\begin{eqnarray}
\nonumber
&&\dfrac{\mathrm{d}}{\mathrm{d}\xi}\mathcal{I}[(y+\xi\eta,y_{x}+\xi\eta_{x},y_{x^2}+\xi\eta_{x^2})]=\\
\nonumber
&&\int\limits_\Omega\left[\dfrac{\partial}{\partial \widehat{y}}\mathcal{F}(\widehat{y},\widehat{y}_{x},\widehat{y}_{x^2},x)-\dfrac{\mathrm{d}}{\mathrm{d}x}\dfrac{\partial}{\partial \widehat{y}_x}\mathcal{F}(\widehat{y},\widehat{y}_{x},\widehat{y}_{x^2},x)+\right.\\
\nonumber
&&\left.\dfrac{\mathrm{d^2}}{\mathrm{d}x^2}\dfrac{\partial}{\partial \widehat{y}_{x^2}}\mathcal{F}(\widehat{y},\widehat{y}_{x},\widehat{y}_{x^2},x)\right]\eta(x)\mathrm{d}x+\\
\nonumber
&&\left[\dfrac{\partial}{\partial \widehat{y}}\mathcal{F}(\widehat{y},\widehat{y}_{x},\widehat{y}_{x^2},x)\eta(x)\right]_{\partial \Omega}+\left[\dfrac{\partial}{\partial \widehat{y}}\mathcal{F}(\widehat{y},\widehat{y}_{x},\widehat{y}_{x^2},x)\eta_x(x)\right]_{\partial \Omega}\\
\label{eq:functional_higher_derivatives_derivate_epsilon_expanded}
&&-\left[\dfrac{\mathrm{d}}{\mathrm{d}x}\dfrac{\partial^2}{\partial \widehat{y}_x^2}\mathcal{F}(\widehat{y},\widehat{y}_{x},\widehat{y}_{x^2},x)\eta(x)\right]_{\partial \Omega}.
\end{eqnarray}
Note that we integrated the last term of Eq.~\eqref{eq:functional_higher_derivatives_derivate_epsilon} twice by parts to obtain $\eta(x)$ under the integral. We obtain a plus sign in front of the third term of the integral. Since $\eta(x)|_{\partial \Omega}=\eta_x(x)|_{\partial \Omega}=0$, the three last terms in Eq.~\eqref{eq:functional_higher_derivatives_derivate_epsilon_expanded} vanish. Finally, since Eq.~\eqref{eq:functional_higher_derivatives_derivate_epsilon_expanded} must be zero for all $\eta(x)$, we obtain, for $\xi=0$
\begin{eqnarray}
\label{eq:ele_second_order}
&&\dfrac{\partial}{\partial y}\mathcal{F}(y,y_{x},y_{x^2},x)-\dfrac{\mathrm{d}}{\mathrm{d}x}\dfrac{\partial}{\partial y_x}\mathcal{F}(y,y_{x},y_{x^2},x)+\dfrac{\mathrm{d^2}}{\mathrm{d}x^2}\dfrac{\partial}{\partial y_{x^2}}\mathcal{F}(y,y_{x},y_{x^2},x)=0.
\end{eqnarray}
This proof can be generalized straightforwardly by imposing that $\eta(x)$ and its $n-1$ derivatives vanish at $\partial \Omega$ for a functional involving $n$ order derivatives. Then we get the Euler-Lagrange for a $n$-order functional
\begin{eqnarray}
\nonumber
&&\dfrac{\partial}{\partial y}\mathcal{F}(y,y_{x},\ldots,y_{x^n},x)+\sum_{k=1}^n(-1)^k\dfrac{\mathrm{d^k}}{\mathrm{d}x^k}\dfrac{\partial}{\partial y_{x^k}}\mathcal{F}(y,y_{x},\ldots,y_{x^n},x)=0
\end{eqnarray}
\end{proof}

\section{Independence of the complex conjugate in the functional}
\label{sec:independence_of_conjugate_functions}

Deriving one function with respect to its complex conjugate in a functional is a \emph{confusing} situation that we physicists have to face from time to time. Usually we try to bypass this uncomfortable dilemma by seeking for an alternative way, as in the case of variations which involve the wave function and its complex conjugate. This cases are specially common in path integral, QED~\cite{Peskin1995} or many-body theory~\cite{Meyer2003,Miyagi2013}, among others. Below, we show that in the derivation of the Euler-Lagrange Equations from the functional~\eqref{eq:jtotal} we can consider $\Psi(\Omega,t)$ and ${\Psi}^*(\Omega,t)$ as independent.
\begin{proof}
Let us define $\mathcal{G}(\Phi,\Psi,\chi,\zeta,\varepsilon)$ as
\begin{eqnarray}
\nonumber
\mathcal{G}(\Phi,\Psi,\chi,\zeta,\varepsilon)&=&-i\int\limits_0^{\widehat{T}}\left[\int_\Omega\Phi(\Omega,t)\left(i\frac{\partial}{\partial t}-H(\Omega, \varepsilon)\right)\chi(\Omega,t)\mathrm{d}\Omega\right]\mathrm{d}t\\
\nonumber
&& +i\int\limits_0^{\widehat{T}}\left[\int_\Omega\zeta(\Omega,t)\left(i\frac{\partial}{\partial t}-H(\Omega, \varepsilon)\right)\Psi(\Omega,t)\mathrm{d}\Omega\right]\mathrm{d}t\\
&& -\alpha\int\limits_0^{\widehat{T}} \left[\varepsilon(t)-\varepsilon_\text{ref}(t)\right]^2\mathrm{d}t + \int_\Omega \Phi O\Psi \mathrm{d}\Omega 
\end{eqnarray}

Applying the Euler-Lagrange equations for $\chi,\zeta,\Phi,\Psi$ and $\varepsilon$ we obtain
\begin{eqnarray}
\label{eqn:g_chi}
i\frac{\partial}{\partial t}\Phi(\Omega,t)&+&H(\Omega,\varepsilon)\Phi(\Omega,t)=0\\
\label{eqn:g_zeta}
i\frac{\partial}{\partial t}\Psi(\Omega,t)&-&H(\Omega,\varepsilon)\Psi(\Omega,t)=0\\
\label{eqn:g_phi}
i\frac{\partial}{\partial t}\chi(\Omega,t)&-&H(\Omega,\varepsilon)\chi(\Omega,t)=-i O\Psi(\Omega,t)\\
\label{eqn:g_psi}
i\frac{\partial}{\partial t}\zeta(\Omega,t)&+&H(\Omega,\varepsilon)\zeta(\Omega,t)=-i O\Phi(\Omega,t) \\
\nonumber
\varepsilon(t)&=&\varepsilon_\text{ref}(t)-\frac{1}{2\alpha}\left[-i\int_\Omega\Phi(\Omega,t)\frac{\partial}{\partial \varepsilon}H(\Omega, \varepsilon)\chi(\Omega,t)\mathrm{d}\Omega\right.\\
\label{eqn:g_varepsilon}
 &&\left. +i\int_\Omega\zeta(\Omega,t)\frac{\partial}{\partial \varepsilon}H(\Omega, \varepsilon)\Psi(\Omega,t)\mathrm{d}\Omega\right]
\end{eqnarray}
By inspecting Eqs.~\eqref{eqn:g_chi} and~\eqref{eqn:g_zeta} we check that ${\Phi}^*(\Omega,t)$ and $\Psi(\Omega,t)$ fulfil the same equation. Furthermore, if we set $\Phi(\Omega,0)=\beta{\Psi}^*(\Omega,0)$, we find that $\Phi(\Omega,t)=\beta{\Psi}^*(\Omega,t)$, being $\beta\in \mathbb{C}$. Analogously, we get that $\zeta(\Omega,t)=\xi{\chi}^*(\Omega,t)$, where $\xi\in\mathbb{C}$. Next, we plug this in Eq.~\eqref{eqn:g_varepsilon}
\begin{eqnarray}
\nonumber
    \varepsilon(t)&=&\varepsilon_\text{ref}(t)-\frac{1}{2\alpha}\left[\beta\int_\Omega{\Psi}^*(\Omega,t)\frac{\partial}{\partial \varepsilon}H(\Omega, \varepsilon)\chi(\Omega,t)\mathrm{d}\Omega\right.\\
    &&\left.+\xi\int_\Omega{\chi}^*(\Omega,t)\frac{\partial}{\partial \varepsilon}H(\Omega, \varepsilon)\Psi(\Omega,t)\mathrm{d}\Omega\right].
\end{eqnarray}
Note that for $\beta=\xi=1$, we obtain
\begin{eqnarray}
&&\left(i\dfrac{\partial}{\partial t}-H(\Omega,\varepsilon)\right)
\psi(\Omega,t)=0,\\
&&\left(i\dfrac{\partial}{\partial t}-H(\Omega,\varepsilon)\right)
\chi(\Omega,t)=-iO\psi(\Omega,t)\delta(t-T),\\
 &&\varepsilon(t)=\varepsilon_\textup{ref}(t)+\frac{1}{\alpha}\operatorname{Im}\int_\Omega\chi^*(\Omega,t)\frac{\partial}{\partial\varepsilon}H(\Omega, \varepsilon)\psi(\Omega,t)\mathrm{d}\Omega,
\end{eqnarray}
and their complex conjugates. 

Summing up, by setting $\Phi(\Omega,0)=\Psi^*(\Omega,0)$ and $\zeta(\Omega,0)=\chi^*(\Omega,0)$ we may derive the QOCT equations~\eqref{eq:psi_eom}-\eqref{eq:epsilon_eom} as done in Sec.~\ref{sec:optimal_control} by assuming that $\Psi(\Omega,t)$ and $\chi(\Omega,t)$ and their complex conjugates are independent.
\end{proof}

\bibliographystyle{unsrt} 
\bibliography{quantum_control_and_information}
\end{document}